# Information Cascades and the Distribution of Economic Recessions in the United States

Paul Ormerod

Volterra Consulting

October 2003

pormerod@volterra.co.uk

*Abstract*

*The phenomenon of information cascades is widely observed in economic and social systems. In a highly connected random network of interacting agents whose decisions are determined by the actions of their neighbours according to a simple threshold rule, the size distribution of cascades is bimodal. There is an exponential tail at small cascade size and a second peak at the size of the entire system corresponding to a single global cascade.*

*The American economy can be thought of as a highly connected network in terms of both its technological and informational connections. The cumulative size of economic recessions, the fall in output from peak to trough, is analysed for the US economy 1900-2002. A least squares fit of an exponential relationship between size and the rank of size gives a good description of most of the data, and technically the null hypothesis that the cumulative size of recessions follows an exponential distribution cannot be rejected at conventional levels of significance. But the observation for the Great Depression of the 1930s does stand out as a very distinct outlier. In other words, we observe evidence of a bimodal relationship of the type predicted by theory.*

## 1.  Introduction

The phenomenon of information cascades [1] is widely observed in economic and social systems.  Ref. [2] notes that small initial shocks can cascade to affect or disrupt large systems that have proved stable with respect to similar disturbances in the past. Examples include financial markets [3], the commercial success or failure of films [4], and the diffusion of crime [5].

Watts [2] offers a general theoretical model of a possible explanation in terms of a sparse, random network of interacting agents whose decisions are determined by the actions of their neighbours according to a simple threshold rule.  Two regimes are identified in which the network is susceptible to very large cascades that occur very rarely.  When cascade propagation is limited by the connectivity of the network, a power law distribution of cascades is observed.  But when the network is highly connected, the size distribution of cascades is bimodal, with an exponential tail at small cascade size and a second peak at the size of the entire system corresponding to a single global cascade.

The purpose of this short paper is to present evidence on the size of recessions in the United States economy, which is an obvious example of a networked system.

## 2.  The background

Firms in the economy are connected in two distinct ways.  First, there are the technological connections which arise from the need for companies in any particular sector to use as inputs into their production the outputs of other industries.  An example is the motor vehicle industry, which uses materials produced by other industries to make cars and trucks.  Input-output tables in the national economic accounts describe these connections at the level of the industry[1]. Ref [6] shows that in a modern developed economy, the connections between industries are rather extensive.

---

[1] similar data is not available at the level of the firm

The second type of connection arises from the impact on any given firm of the decisions and opinions about the future of other firms in the economy. Keynes [7], for example, attached considerable importance to the state of long term expectations for the decisions by companies to invest. The more optimistic the overall climate of opinion, the more likely it is that a firm will carry out an investment decision. Ref [8] sets out an agent based theoretical model based on this principle which is able to generate a good approximation to some of the key underlying properties of US business cycle time series data.

Information about the general climate of expectations is readily available from the media, through newspapers such as the *Wall Street Journal*, for example. This is certainly the case as far as large companies are concerned. In other words, it is as if all agents are connected to, say, Microsoft or General Motors in terms of gathering information about the business climate. The information they gather may not be perfect, but a large amount of it is available.

So on both the technological and informational criteria, the US economy appears to be highly connected.

The duration of recessions in 17 capitalist economies over the 1871-1994 period is examined in [9], a recession being defined as a period of at least one year in which the rate of growth of real GDP was less than zero. The individual country data on duration was pooled, giving a total number of 336 observations across the 17 country sample.

A power law relationship between frequency and duration of recessions provides a good approximation to the data, but suggests that there should have been more recessions of long duration than have actually existed. However, when recessions lasting only one year are excluded from the sample, a power law relationship with an exponent of $-3.2$ provides an almost perfect fit to the data. Ref [9] concludes that 'there may be two separate processes going on in the process which generates data on capitalist recessions. When a recession arises, for whatever reason, agents appear to have some capacity to react quickly, which often prevents the recession from being prolonged beyond one year. Once this has happened, however, recessions can take place on all scales of duration'. In other words, there appears to be a bimodal type of relationship in the duration of recessions in capitalist economies.

Ref [9] fitted power law relationships to the cumulative size of recessions, and although such a relationship provides a rough approximation to the data, it is much less satisfactory than the duration relationship.

*3.     The results*

Percentage changes in real annual GDP are analysed for the United States over the period 1900-2002. The data source from 1929 onwards is [10] and before that date [11].

The cumulative size of recessions is the variable of interest, in other words the percentage reduction in the level of GDP from peak to trough. There are in total 19 observations, ranging in size from the fall of barely one fifth of one per cent in GDP in the recession of 1982, to the fall of some 30 per cent in the Great Depression of 1930-33. Figure 1 plots the histogram of the data, using absolute values.

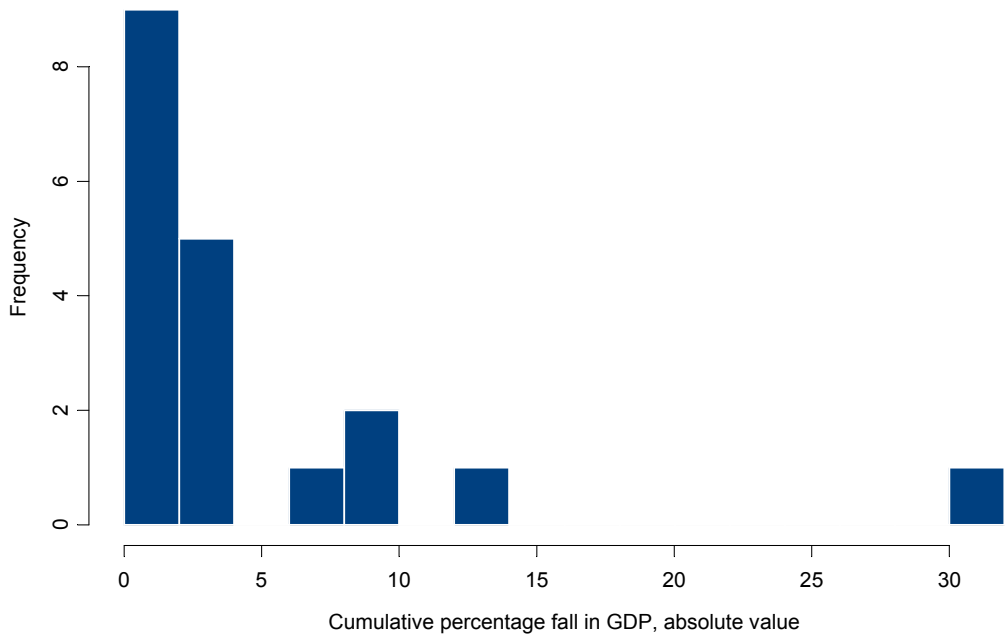

**Figure 1** *Histogram of cumulative absolute percentage fall in real US GDP, peak to trough, during recessions 1900-2002*

Most recessions have been fairly small. By far the largest recession was the Great Depression of the early 1930s, and the second largest was the transition back into a peace-time economy at the end of World War Two.

The cumulative density function is plotted in Figure 2, along with that of an exponential distribution with rate = 0.3.

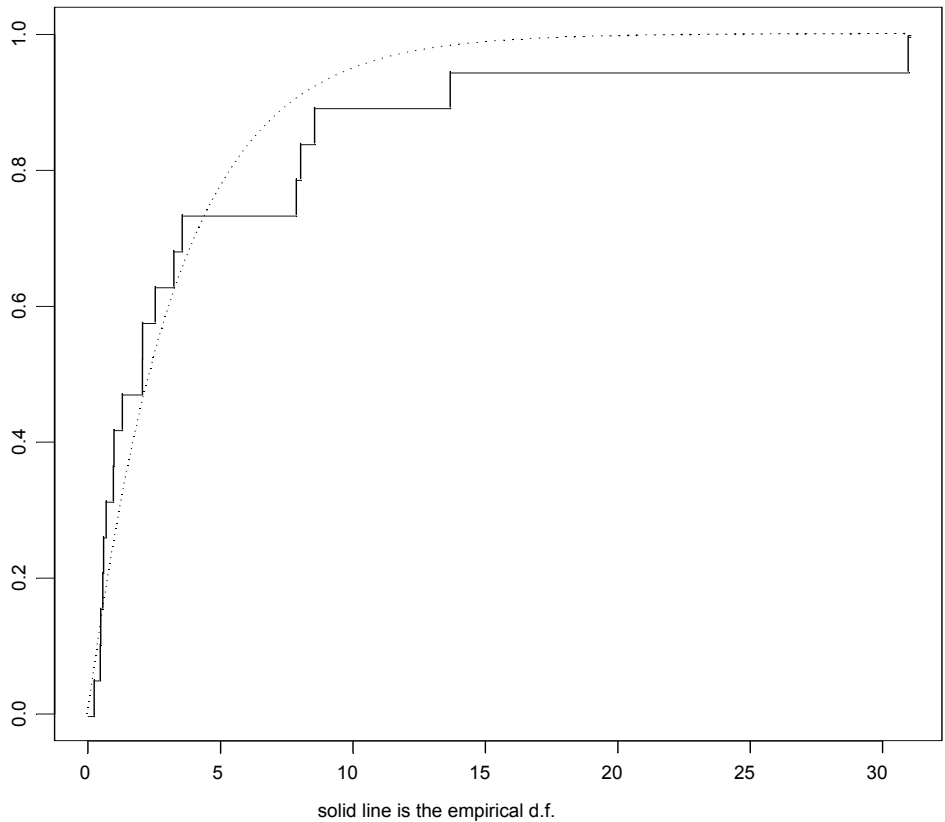

**Figure 2**  *Cumulative density function of cumulative absolute percentage fall in real US GDP, peak to trough, during recessions 1900-2002. Solid line is the empirical d.f and the dotted line is that of an exponential distribution with rate = 0.3*

Technically, on a Kolmogorov-Smirnov test, the null hypothesis that the data are distributed exponentially with rate = 0.3 is not rejected at the standard levels of statistical significance.

A formal regression shows that an exponential relationship between size and rank gives a good description of the data. The fit is distinctly better than that of a power law, with the regression standard error being 0.347 for the power law and 0.252 for the exponential.

However, the outlying observation of the Great Depression period is predicted very inaccurately by the exponential fit. Table 1 shows the actual values of the six largest cumulative recessions, and the values fitted by the exponential relationship between size and rank.  The smaller recessions are fitted very closely by the relationship.

**Table 1    Six largest cumulative recessions, percentage change in real GDP, actual and fitted from exponential least-squares regression of cumulative size on rank**

| | | | | | | |
|---|---|---|---|---|---|---|
| **Actual** | 3.6 | 7.9 | 8.0 | 8.6 | 13.7 | 31.0 |
| **Fitted** | 5.0 | 6.3 | 8.0 | 10.1 | 12.7 | 16.1 |

The overall fit to the tail is not perfect, but the Great Depression clearly stands out as an outlier.

Excluding the observation for the Great Depression makes little difference to the estimated coefficient on the rank of the cumulative size.  Using the whole sample it is 0.234 with a standard error of 0.011 and excluding the 1930s depression it is 0.221 with a standard error of 0.008.  But the fit to the data is good, with the multiple $R^2$ being 0.978 excluding the Great Depression observation.

In other words, the bulk of the data is fitted well by an exponential relationship between size and rank of size, but there is evidence of a bimodal distribution with the observation for the Great Depression of the 1930s being a clear outlier.

The American economy is a highly connected network.  In such circumstances, when it is as if the degree of connectivity between agents is high, ref [2] suggests that a bimodal distribution of cascades will be observed, with an exponential tail at small cascade size, and a second peak with a much larger, global cascade size.  This appears to be very close to what we actually observe.

*4.    Conclusion*

In a random network of interacting agents whose decisions are determined by the actions of their neighbours according to a simple threshold rule, [2] suggests that when the network is highly connected, a bimodal distribution of cascades will be observed. There will be an exponential tail at small cascade size and a second peak at the size of the entire system corresponding to a single global cascade.

The American economy can be thought of as a highly connected network, in terms of both the technological connections between industries and the access to information on the decisions and opinions of agents about economic prospects.

In such circumstances, when it is as if the degree of connectivity between agents is high, ref [2] suggests that a bimodal distribution of cascades will be observed, with an exponential tail at small cascade size, and a second peak with a much larger, global cascade size. This appears to be very close to what we actually observe.

The cumulative percentage fall in real GDP from peak to trough in American recessions 1900 – 2002 can be approximated closely by a bimodal distribution. An exponential relationship between size and the rank of size fits the bulk of the data well, but the single observation of the Great Depression of the 1930s stands out as an outlier.